\documentclass[aps,floatfix,superscriptaddress]{revtex4}
\usepackage{amssymb,amsmath,amsfonts,latexsym,graphicx,epsfig,here,color}
\newcommand{\bea}{\begin{eqnarray}}
\newcommand{\ena}{\end{eqnarray}}
\newcommand{\be}{\begin{equation}}
\newcommand{\en}{\end{equation}}
\newcommand{\nn}{\nonumber\\}

\newcommand{\la}{\langle}
\newcommand{\ra}{\rangle}

\begin{document}

\hfill MITP/18-068 (Mainz)

\title{Analysis of the semileptonic and nonleptonic two-body decays of \\
the double heavy charm baryon states $\Xi_{cc}^{++}, \,\Xi_{cc}^{+}$ and 
$\Omega_{cc}^+$} 

\author{Thomas Gutsche}
\affiliation{Institut f\"ur Theoretische Physik, Universit\"at T\"ubingen,
Kepler Center for Astro and Particle Physics, 
Auf der Morgenstelle 14, D-72076 T\"ubingen, Germany}
\author{Mikhail~A.~Ivanov}
\affiliation{Bogoliubov Laboratory of Theoretical Physics,
Joint Institute for Nuclear Research, 141980 Dubna, Russia}
\author{J\"urgen~G.~K\"orner}
\affiliation{PRISMA Cluster of Excellence, Institut f\"{u}r Physik,
Johannes Gutenberg-Universit\"{a}t, D-55099 Mainz, Germany}
\author{Valery E. Lyubovitskij}
\affiliation{Institut f\"ur Theoretische Physik, Universit\"at T\"ubingen,
Kepler Center for Astro and Particle Physics,
Auf der Morgenstelle 14, D-72076 T\"ubingen, Germany}
\affiliation{Departamento de F\'\i sica y Centro Cient\'\i fico
Tecnol\'ogico de Valpara\'\i so-CCTVal, Universidad T\'ecnica
Federico Santa Mar\'\i a, Casilla 110-V, Valpara\'\i so, Chile}
\affiliation{Department of Physics, Tomsk State University,
634050 Tomsk, Russia} 
\author{Zhomart Tyulemissov}
\affiliation{Bogoliubov Laboratory of Theoretical Physics,
Joint Institute for Nuclear Research, 141980 Dubna, Russia}
\affiliation{The Institute of Nuclear Physics, Ministry of Energy of 
the Republic of Kazakhstan, 050032 Almaty, Kazakhstan}
\affiliation{Al-Farabi Kazakh National University, 050038 Almaty, 
Kazakhstan}

\begin{abstract}

We calculate the semileptonic and a subclass of sixteen nonleptonic two-body 
decays of the double charm baryon ground states 
$\Xi_{cc}^{++},\,\Xi_{cc}^{+}$ and $\Omega_{cc}^+$ where we concentrate 
on the nonleptonic decay modes. We identify those nonleptonic decay channels 
in which the decay proceeds solely via the factorizing contribution 
precluding a contamination from $W$ exchange. We use the covariant confined 
quark model previously developed by us to calculate the various helicity 
amplitudes which describe the dynamics of the $1/2^+ \to 1/2^+$ and 
$1/2^+ \to 3/2^+$ transitions induced by the Cabibbo-favored effective 
$(c \to s)$ and $(d \to u)$ currents. We then proceed to calculate 
the rates of the decays as well as polarization effects and angular decay 
distributions of the prominent decay chains resulting from the nonleptonic 
decays of the double heavy charm baryon parent states.

\end{abstract}

\today

\maketitle

\section{Introduction}

Two years ago the LHCb Collaboration reported on the discovery of the
double heavy charm baryon state $\Xi_{cc}^{++}$~\cite{Aaij:2017ueg}. 
The state was found in the invariant mass spectrum of the final state 
particles $(\Lambda_c^+\,K^-\,\pi^+\,\pi^+)$ where the $\Lambda_c^+$ baryon 
was reconstructed in the decay mode $pK^-\pi^+$. The mass of the new state 
was given as $3621.40 \pm0.72 \pm0.14 \pm 0.27$ MeV.
A year later the LHCb Collaboration identified the same state in the decay 
$\Xi_{cc}^{++} \to \Xi_c^+ +\pi^+$ with a mass value of
$3620.6 \pm1.5(\rm{stat})\pm 0.4(\rm{syst})\pm 
0.3(\Xi_c^+)$ MeV~\cite{Aaij:2018gfl}. The lifetime of the $\Xi_{cc}^{++}$ 
was measured to be $\tau(\Xi_{cc}^{++})=
0.256^{+0.024}_{-0.022}(\rm{stat})\pm 0.014(\rm{syst})$ 
ps~\cite{Aaij:2018wzf}.

The weighted average of the two mass measurements of 
$m_{\Xi_{cc}^{++}}= 3621 \pm 1.1 \pm 0.3 \pm 0.3$ MeV is
quite close to the value of 3610 MeV predicted 
some time ago in Ref.~\cite{Korner:1994nh} in the framework of the one-gluon
exchange model of de Rujula {\it et al.}~\cite{DeRujula:1975smg} with 
a Breit-Fermi spin-spin interaction term. It is noteworthy
that Ebert {\it et al.} 
predicted a mass of 3620 MeV for the $\Xi_{cc}^{++}$ using 
a relativistic quark-diquark potential model~\cite{Ebert:2002ig}.
In Ref.~\cite{Gutsche:2017hux} we have interpreted the new double charm baryon
state found in the $(\Lambda_c^+\,K^-\,\pi^+\,\pi^+)$ mass distribution
as being at the origin of the decay chain 
$\Xi_{cc}^{++} \to \Sigma_c^{++}(2455;1/2^+) (\to \Lambda_c^+ \pi^+)
+ \bar K^{*0} (\to K^-  \pi^+)$. In the present paper we extend the analysis 
of Ref.~\cite{Gutsche:2017hux} in two directions. First, we consider the 
possibility that the first step in the decay chain consists of the decay
$\Xi_{cc}^{++} \to \Sigma_c^{++}(2520;3/2^+) + \bar K^{*0}$ where the state 
$\Sigma_c^{++}(2520;3/2^+)$ is the spin $3/2$ heavy quark symmetry partner 
of the $\Sigma_c^{++}(2455;1/2^+)$. In fact, in a talk at a CERN Seminar~\cite{Chang17} 
Zhang (LHCb Collaboration) showed 
an invariant mass plot for the $(\Lambda_c \,\pi^+)$ subsystem in which 
the peaking bin for $m(\Lambda_c \pi^+)$ lies in between the two 
$\Sigma_c^{++}(2455;1/2^+)$ and $\Sigma_c^{++}(2520;3/2^+)$ states.
Second, we provide results for a subclass of the Cabibbo-favored nonleptonic 
two-body decays of the not yet identified $J^P=1/2^+$ double charm
baryon ground states $\Xi_{cc}^{+}(3610)$ and $\Omega_{cc}^+(3710)$ where the
mass values are again taken from the calculation of~\cite{Korner:1994nh}.
The authors of~\cite{Ebert:2002ig}
predict a mass value of $M_{\Omega_{cc}^+}=3778$ MeV which is considerably 
higher than the value $M_{\Omega_{cc}^+}=3710$ calculated 
in Ref.~\cite{Korner:1994nh}. A recent lattice calculation quotes a value of 
$3712 \pm 10 \pm 12$ MeV for the $\Omega_{cc}^+$ state~\cite{Mathur:2018rwu}.

The physics of double heavy charm and bottom baryons (mass spectrum and 
decay properties) has been studied before in a number of  
papers~\cite{Korner:1994nh,DeRujula:1975smg,Ebert:2002ig,Gutsche:2017hux},  
\cite{Mathur:2018rwu}-\cite{Shi:2019fph}. 
We presented a detailed analysis of exclusive decays of double heavy baryons 
using several versions of covariant quark models in 
Refs.~\cite{Faessler:2001mr,Gutsche:2017hux,Gutsche:2019wgu}.
Double heavy baryon decays and their magnetic moments were treated by us 
in Refs.~\cite{Faessler:2001mr} where we performed a comprehensive 
study of the semileptonic and radiative decays of double heavy baryons using 
a covariant quark model without implementing quark confinement. 
The version of the covariant quark 
model used in~\cite{Faessler:2001mr} has been improved by 
incorporating quark confinement in an effective way~\cite{Branz:2009cd}. 
For the calculation of the relevant $1/2^+ \to 1/2^+$ and $1/2^+ \to 3/2^+$ 
transitions performed in this paper we use the improved quark model which 
we refer to as  the covariant confined quark model (CCQM). 
In Refs.~\cite{Gutsche:2017hux,Gutsche:2018msz,Gutsche:2019wgu} we studied 
decay properties of double heavy baryons in the CCQM approach. 
In particular, in Ref.~\cite{Gutsche:2017hux} we interpreted the 
$\Xi_{cc}^{++}$ baryon found by the LHCb Collaboration in the invariant 
mass distribution of the set of final state particles 
$(\Lambda^+_c K^− \pi^+ \pi^+)$ as being at the origin of the decay chain 
$\Xi_{cc}^{++} \to \Sigma_c^{++} (\to \Lambda_c^+ \pi^+)
+\bar K^{*0}(\to K^- \pi^+)$. The nonleptonic decay $\Xi_{cc}^{++} \to 
\Sigma_c^{++} \bar K^{*0}$ belongs to the class of factorizing decays, 
i.e. the decays precluding a contamination from internal $W$ exchange. 
As a byproduct of our investigation we have also analyzed the nonleptonic 
mode with $\bar K^0$ in the final state. In Ref.~\cite{Gutsche:2018msz} 
we proposed a novel method for the evaluation of the nonfactorizing 
(three quark loop) diagrams generated by $W$--exchange and 
contributing to the nonleptonic two-body decays of the
doubly charmed baryons $\Xi_{cc}^{++}$ and $\Omega_{cc}^{+}$. 
The $W$--exchange contributions appear in addition to the factorizable
tree graph contributions and are not suppressed in general. 
In Ref.~\cite{Gutsche:2019wgu} we reviewed novel ideas in the  
theoretical description of nonleptonic decays of double heavy baryons. 
In the present paper we extend our analysis of semileptonic 
decays of double charm baryons started in Ref.~\cite{Faessler:2001mr} 
by inclusion of all factorizable modes for both types of 
weak transitions --- semileptonic and nonleptonic using 
the updated theoretical framework --- CCQM model. 
Note that in our paper Ref.~\cite{Faessler:2001mr} we used mass  
values for the single and double charm baryons masses 
$\Xi_{cc}^{++} = 3.61$ GeV and $\Xi_c^{' +} = 2.47$ GeV,  
which differ from the updated mass values used in the present paper
($\Xi_{cc}^{++} = 3.6206$ GeV and $\Xi_c^{' +} = 2.5774$ 
GeV~\cite{Tanabashi:2018oca}). When comparing the relevant semileptonic rate
in~\cite{Faessler:2001mr} to that in the present paper one has to take the
changed mass values into account, which results in a suppression of our 2001
result by a factor of $\sim 1.6$ which is mostly kinematical in nature. 

Our paper is structured as follows. In Sec.~II we discuss the 
decay topologies of the Cabibbo-favored nonleptonic two-body decays 
of the double charm baryon ground states $\Xi_{cc}^{++},\,\Xi_{cc}^{+}$ 
and $\Omega_{cc}^+$. Of the many possible decays we identify 16 decays 
which proceed via the factorizing contributions alone. In Sec.~III 
we collect material on the spin kinematics of the decays. We define 
invariant form factors and helicity amplitudes. We also write down 
formulas for the semileptonic and nonleptonic rates. In Sec.~IV we list 
a set of local interpolating three-quark currents with the correct 
quantum numbers of the baryon states that they describe. The nonlocal 
versions of the interpolating currents enter the calculation of  
the various transition form factors in our covariant confined quark 
model (CCQM). We also give a brief description of the main features 
of our CCQM calculation. Sec.~V contains our numerical results for the 
semileptonic and nonleptonic rates and branching fractions. In Sec.~V  
we also discuss polarization effects and angular decay distributions  
of the eight semileptonic and 16 nonleptonic cascade decays. In Sec.~VI 
we summarize our results and outline our follow-up program of further 
calculations involving also $W$--exchange contributions to the 
nonleptonic decays of double heavy charm baryons. 
  
\section{Decay topologies of Cabibbo-favored double heavy charm baryon
nonleptonic decays}

\begin{figure}[htb]
\begin{center}
\epsfig{figure=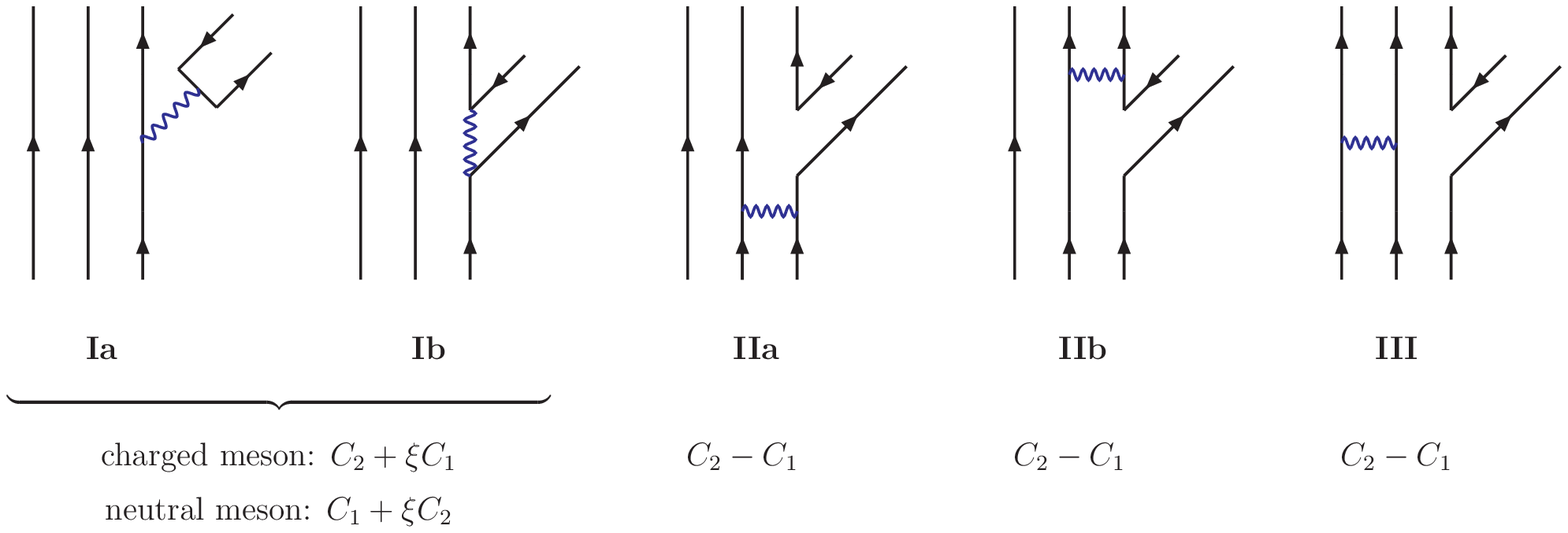,scale=.8}
\caption{Flavor-color topologies of nonleptonic weak decays.}
\label{fig:NLWD}
\end{center}
\end{figure}

We begin by a discussion of the different color-flavor topologies that 
contribute to the various possible nonleptonic two-body transitions of the 
double heavy $\Xi_{cc}^{++},\,\Xi_{cc}^{+}$ and $\Omega_{cc}^+$ states. 
The relevant topologies 
are displayed in Fig.~\ref{fig:NLWD}. We refer to the topologies of Ia 
and Ib as tree diagrams. They are sometimes also 
referred to as external (Ia) and internal $W$-emission (Ib) diagrams. 
The topologies IIa, IIb and III are referred to as $W$--exchange diagrams.  
In \cite{Leibovich:2003tw} they are denoted as the Exchange (IIa), 
color-commensurate (IIb) and Bow tie (III) diagram. As shown in 
Fig.~\ref{fig:NLWD} the color-flavor factor of the tree diagrams Ia and Ib 
depends on whether the emitted meson is charged or neutral. For charged 
emission the color-flavor factor is given by the linear combination 
of Wilson coefficients $(C_2 + \xi C_1)$, where 
$\xi=1/N_c$, while for neutral emission the color-flavor factor reads 
$(C_1 + \xi C_2)$. We take $C_1=-0.51$ and $C_2=1.20$ from 
Ref.~\cite{Buchalla:1995vs}. We use the large $N_c$ limit for the 
color-flavor factors. For the $W$--exchange diagrams not treated in this 
paper the color-flavor factor is given by $(C_2-C_1)$.

In Table~\ref{topol} we provide a complete list of the Cabibbo-favored 
ground-state to ground-state nonleptonic two-body decays of double heavy 
charm baryons together with the color-flavor topologies that contribute 
to these decays. For reasons of compactness we employ a star notation for 
the spin 3/2 ground state baryons which differs from the notation 
suggested by the Particle Data Group (PDG). 
Thus, for example, our $\Omega_c^{*\,0}$ 
stands for the spin 3/2 partner of the spin 1/2 state $\Omega_c^{0}$. 
The spin 3/2 state $\Omega_c^{*\,0}$ is listed in 
the PDG~\cite{Tanabashi:2018oca} as $\Omega_c(2770)^0$.
   
\begin{table}[htb]
\caption
    {Cabibbo-favored nonleptonic two-body decays of double
      heavy charm baryons including $W$--exchange contributions.}
   \vspace{0.2cm}
\begin{center}
\begin{tabular}{lccccc}
  \hline
  \hline
\phantom{$1/2^+ \to 1/2^+ +0^-$}\qquad& $I_a$ &$I_b$ &
$II_a$ & $II_b$ & $III$ \\ 
      \hline
$\Xi_{cc}^{++} \to \Sigma_{c}^{(*)\,++} + \bar K^{(*)\,0}$\quad \quad \quad
      &-
      \quad \quad&$\surd$ \quad \quad& $-$\quad \quad& $-$ \quad \quad&$-$
      \\[0.5ex]
$\Xi_{cc}^{++} \to \Xi_{c}^{(\prime,* )\,+} + \pi^+(\rho^+)$\quad \quad \quad
      &$\surd$
      \quad \quad&$-$ \quad \quad& $-$\quad \quad& $\surd$\quad \quad&$-$
      \\[0.5ex]
$\Xi_{cc}^{++} \to \Sigma^{(*)\,+} + D^{(*)+}$\quad \quad \quad &$-$
      \quad \quad &$-$ \quad \quad & $-$\quad \quad& $\surd$\quad \quad&$-$
      \\[0.5ex]
$\Xi^+_{cc} \to \Xi_{c}^{(\prime ,*)\,0}+ \pi^+(\rho^+)$ \quad \quad \quad & 
$\surd$\quad \quad &$-$\quad \quad &$\surd $\quad \quad &$-$\quad \quad &$-$
 \\[0.5ex]
$\Xi_{cc}^{+} \to \Lambda_c^{+}(\Sigma_c^{(*)+})
 +\bar K^{(*)0}$ &$-$ &$\surd$&$\surd$ & 
$-$ &$-$\\[0.5ex]
$\Xi_{cc}^{+} \to \Sigma_c^{(*)++}
 + K^{(*)-}$ &$-$ &$-$ & 
 $\surd$ &$-$&$-$\\[0.5ex]
 $\Xi^+_{cc} \to \Xi_{c}^{(\prime ,*)\,+}+ \pi^0(\rho^0)$
&$-$ &$-$ & $\surd$ &$\surd$&$-$\\[0.5ex]
 $\Xi^+_{cc} \to \Xi_c^{(\prime ,*)\,+}+ \eta(\eta^\prime )$
 &$-$ &$-$ &$\surd$ &$\surd$&$-$\\[0.5ex]
 $\Xi^+_{cc} \to \Omega_c^{(*)\,0} + K^{(*)\,+}$
&$-$ &$-$ &$\surd$ &$-$&$-$\\[0.5ex]
 $\Xi_{cc}^{+} \to \Lambda^{0}(\Sigma^{(*)0})
 + D^{(*)+}$ &$-$& $-$ &$-$ &$\surd$ &$\surd$
 \\[0.5ex]
 $\Xi_{cc}^{+} \to \Sigma^{(*)+}+ D^{(*)0}$
 &$-$ &$-$ &$-$& $-$ &$\surd$\\[0.5ex]
 $\Xi_{cc}^{+} \to \Xi^{(*)0}+ D_s^{(*)+}$
 &$-$ &$-$ &$-$& $-$ &$\surd$\\[0.5ex]
 $\Omega_{cc}^{+} \to \Xi_{c}^{(\prime ,*)\,+} 
 +\bar {K}^{(*)0}$&$-$ &$\surd$ &$-$ &$\surd$ &$-$ \\[0.5ex]
 $\Omega_{cc}^{+} \to \Xi^{(*)\,0} 
 + D^{(*)+}$ &$-$ &$-$ &$-$ &$\surd$ &$-$ \\[0.5ex]
 $\Omega_{cc}^{+} \to \Omega_c^{(*)\,0} 
 + \pi^+(\rho^+)$ &$\surd$ &$-$ &$-$ &$-$ &$-$ \\[0.5ex]
\hline
\hline
\end{tabular} 
\end{center}
\label{topol}
\end{table}
     
In this paper we restrict our analysis to those nonleptonic decays 
whose decay dynamics is solely determined by the tree diagram contributions 
Ia and Ib. There are two classes of such decays which we discuss in turn.
\begin{itemize}
\item The first class of decays is solely contributed to by the two 
topologies Ia and Ib.  
These decays can be identified by necessary and sufficient
conditions for the quarks involved in the two-body nonleptonic transitions 
which we label according to the following scheme
\be
B_1(q_1\,q_2\,q_3) \to B_2(q'_1\,q'_2\,q'_3) + M(q_m \bar q_{\bar n})\,.
\en
A necessary condition for the contribution of the factorizing class 
of decays is that a quark pair $q_{i}q_{j}=q'_{i}q'_{j}$
is shared by the parent and daughter baryon $B_{1}$ and $B_{2}$, respectively.
A sufficient condition for the factorizing class of decays is that
(i) $ q_m$ is not among $q_1,\,q_2,\,q_3$ and
(ii) $q_{\bar n}$ is not among $q'_1,\,q'_2,\,q'_3$. 
Using these two criteria we have identified the two groups of decays
\be
\label{class1}
\Xi_{cc}^{++} \to \Sigma_{c}^{(*)++} + \bar K^{(*)0 } \qquad \qquad 
\Omega_{cc}^{+} \to \Omega_c^{(*)0} + \pi^+(\rho^+)
\en
which proceed via the tree graphs alone.
\item The second class of decays involves in addition to the 
tree topologies also the $W$--exchange topologies IIb which,  
however, do not contribute because of the K\"orner, Pati, and Woo (KPW) 
theorem \cite{Korner:1970xq,Pati:1970fg}. The K\"orner, Pati, and Woo 
theorem states that the contraction of the  
flavor antisymmetric current-current operator with a flavor symmetric 
final state configuration is 0. There are two groups of decays that 
belong to this class given by
\be
\label{class2}
   \Xi_{cc}^{++} \to \Xi_{c}^{\prime(*)+} + \pi^+(\rho^+) \qquad \qquad
   \Omega_{cc}^{+} \to \Xi_{c}^{\prime (*)+} 
   +\bar {K}^{(*)0}
\en 
We neglect SU(3) breaking effects when applying the KPW theorem to the above 
two groups of decays. We plan to quantify the SU(3) breaking effects 
 in a future dynamical calculation of these decays. 
\end{itemize} 
The recently observed decay 
$\Xi_{cc}^{++} \to \Xi_c^+ +\pi^+$~\cite{Aaij:2018gfl} is not discussed
in this paper since, in addition to the tree diagram Ia, there 
is a non-vanishing contribution from the $W$--exchange diagram IIb. 
In this context it is interesting to observe that the decays
\bea
 &&\Xi_{cc}^{++} \to \Sigma^{\ast \,+} + D^{(*)+}  \qquad
 \Xi_{cc}^{+} \to \Sigma^{*0}+ D^{(*)+} \qquad 
 \Xi_{cc}^{+} \to \Sigma^{*+}+ D^{(*)0}   \nn
 && \Xi_{cc}^{+} \to \Xi^{*0}+ D_s^{(*)+}   \qquad
\quad  \Omega_{cc}^{+} \to \Xi^{\prime (*)0} 
 + D^{(*)+}
\ena
induced by the topologies IIb and III are predicted to be altogether zero 
due to the KPW theorem. It would be very interesting to experimentally 
confirm this prediction.

Let us add a few comments concerning the $W$--exchange diagrams. 
The contribution of the $W$--exchange diagrams 
cannot be neglected even if this is frequently done in the analysis 
of nonleptonic charm baryon decays. A prominent example is the decay 
$\Lambda_c^+ \to \Sigma^0 \pi^+$ which is contributed to by the topologies 
IIa, IIb, and III, i.e. there are no tree graph contributions to the decay. 
Nevertheless, its experimental branching ratio is comparable to that of the 
decay mode $\Lambda_c^+\to \Lambda^0 \pi^+$ where the latter mode is also 
contributed to by the tree diagram 1a. The interplay of the tree and 
$W$--exchange diagrams for the Cabibbo-favored $\Delta C=1$ nonleptonic 
charm baryon decays has been studied in~\cite{Korner:1978tc,%
Korner:1992wi,Uppal:1994pt,Fayyazuddin:1996iy} and also in a previous 
version of our model~\cite{Ivanov:1997ra}. We hope to return to the
calculation of the $W$--exchange contributions in single 
charm and double charm baryon decays in the framework of our CCQM quark 
model. We mention that the evaluation of the $W$-exchange diagrams in our 
approach is technically quite demanding since it involves 
a three-loop calculation. Naturally it is of utmost importance to get the 
relative signs between the tree and $W$--exchange contributions right since 
this decides whether the two classes of contributions interfere 
constructively or destructively. A first attempt to estimate 
the $W$--exchange contributions to the $1/2^+ \to 1/2^+ + 0^-$ double heavy 
baryon decays has been published in~\cite{Sharma:2017txj} using a baryon pole 
model for the $W$--exchange contributions. There is a one-to-one 
correspondence between the $1/2^+ \to 1/2^+ + 0^-$ decays treated 
in~\cite{Sharma:2017txj} and the $1/2^+ \to 1/2^+ + 0^-$ decays listed in 
Table~\ref{topol}. 
The $W$--exchange topology structure of the decays written down 
in~\cite{Sharma:2017txj} in terms of $s$--channel and $u$--channel 
contributions is consistent with the corresponding topology structure in 
Table~\ref{topol}.

Returning to the factorizing contributions we in the following 
discuss the class 1 and class 2 decays listed in Eqs.~(\ref{class1}) 
and~(\ref{class2}), respectively, which are determined by the 
factorizing contributions alone.
   
\section{Matrix elements, helicity amplitudes and decay rates expressions}

The matrix element of the exclusive decay 
$B_1(p_1,\lambda_1)\to B_2(p_2,\lambda_2)\,+\,M(q,\lambda_M)$ 
is defined by ($p_1=p_2+q$)
\be
M(B_1\to B_2 + M) =  
\frac{G_F}{\sqrt{2}} \, V_{ij} \, V^\ast_{kl} \, C_{\rm eff} \, 
f_M \, M_M \, \la B_2 | \bar q_2 O_\mu q_1 | B_1 \ra \, 
\epsilon^{\dagger\,\mu}(\lambda_M) \,,
\label{eq:matr_LbLJ}
\en 
where $M=V$ and $M=P$ stand for the vector and pseudoscalar meson cases  
such that $M_M$ and $f_M$ are the respective masses $M_V,\,M_P$ and leptonic 
decay constants $f_V,\,f_P$.  The Dirac string $O^\mu$ is defined by
$O^\mu = \gamma^\mu (1 - \gamma^5)$. Here $V_{ij}$ 
are the Cabibbo-Kabayashi-Maskawa (CKM) matrix elements: 
$V_{ud} = 0.97420$ and $V_{cs} = 0.997$. 

Here $C_{\rm eff}$ is the combination of the Wilson coefficients 
$(C_2 + \xi C_1)$, where $\xi=1/N_c$ and $N_c$ is the number of colors,
while for neutral emission the color-flavor factor reads $(C_1 + \xi C_2)$. 
We take  $C_1=-0.51$ and $C_2=1.20$ at
$\mu=m_c=1.3$~GeV from Ref.~\cite{Buchalla:1995vs}. We use the large $N_c$
limit for  the color-flavor factors.
It is known, that nonfactorizable contributions coming from,
e.g., one-gluon exchange, might be important for the description of
nonleptonic decays. As an example, recall the well-known decay
$B \to J/\psi+K$, which is proportional to the coefficient
$a_2 = C_1 + \xi C_2$ and would thus be predicted to be 0 for $N_c=3$.
In this case naive factorization clearly does not describe the experimental
data. The discussion on and determination of 
the nonfactorizable corrections
to the coefficient $a_2$ has been actively pursued in the literature using
various techniques. However, as far as we know, up to now
there is no well-established framework in which the nonfactorizable
contributions could be taken into account in a self-consistent way.
Therefore, we employ a phenomenological and simple assumption
in our calculations of the nonleptonic decays of both heavy mesons and baryons
that the color factor $\xi=1/N_c$ appearing in the combination of the Wilson
coefficients is set to 0. This assumption has been extensively used in the
literature and is well justified in the comparison with experimental data in
the meson sector.

The hadronic matrix element $\la B_2 | \bar q_2 O^\mu q_1 | B_1 \ra$
can be expressed in terms of six ($1/2^+ \to 1/2^+$) and eight
($1/2^+ \to 3/2^+$) dimensionless invariant form factors
$F^{V/A}_i(q^2)$, respectively. One has 

\noindent
for the transition $\frac{1}{2}^+ \to \frac{1}{2}^+$\,: 
\bea 
\label{ff1/2}
\la B_2 | \bar q_2 \gamma_\mu q_1 | B_1 \ra &=& 
\bar u(p_2,s_2) 
\Big[ \gamma_\mu F_1^V(q^2) 
    - i \sigma_{\mu\nu} \frac{q_\nu}{M_1} F_2^V(q^2)  
    + \frac{q_\mu}{M_1} F_3^V(q^2) 
\Big] 
u(p_1,s_1)\,, \nonumber\\ 
\la B_2 |\bar q_2\gamma_\mu\gamma_5 q_1  | B_1 \ra &=& 
\bar u(p_2,s_2) 
\Big[ \gamma_\mu F_1^A(q^2)  
    - i \sigma_{\mu\nu} \frac{q_\nu}{M_1} F_2^A(q^2)   
    + \frac{q_\mu}{M_1} F_3^A(q^2)  
\Big] \gamma_5 u(p_1,s_1) 
\ena
 
\noindent
and for the transition $\frac{1}{2}^+ \to \frac{3}{2}^+$\,: 
\bea
\label{ff3/2}
\la B_2^\ast |\bar q_2 \gamma_\mu q_1| B_1 \ra &=& 
\bar u^\alpha(p_2,s_2) 
\Big[ g_{\alpha\mu} F_1^V(q^2)  
    + \gamma_\mu \frac{p_{1\alpha}}{M_1} F_2^V(q^2) 
+ \frac{p_{1\alpha} p_{2\mu}}{M_1^2} F_3^V(q^2)   
    + \frac{p_{1\alpha} q_\mu}{M_1^2}    F_4^V(q^2)       
\Big] \gamma_5 
u(p_1,s_1) \,, \nonumber\\
\la B_2^\ast |\bar q_2 \gamma_\mu\gamma_5 q_1 | B_1 \ra &=& 
\bar u^\alpha(p_2,s_2) 
\Big[ g_{\alpha\mu} F_1^A(q^2)  
    + \gamma_\mu \frac{p_{1\alpha}}{M_1} F_2^A(q^2) 
+ \frac{p_{1\alpha} p_{2\mu}}{M_1^2} F_3^A(q^2)   
    + \frac{p_{1\alpha} q_\mu}{M_1^2}    F_4^A(q^2)       
\Big] 
u(p_1,s_1) 
\ena  
\noindent where 
$\sigma_{\mu\nu} = (i/2) (\gamma_\mu \gamma_\nu - \gamma_\nu \gamma_\mu)$ 
and all $\gamma$ matrices are defined as in Bjorken-Drell. 

The results of a covariant dynamical calculation as in the present
case are usually obtained in terms of the invariant form factors defined above. 
To proceed further, it
is very convenient to convert the set of invariant form factors to a set
of helicity amplitudes where the two sets are linearly related.
We therefore express the vector and axial helicity amplitudes 
$H^{V/A}_{\lambda_2\lambda_M}$ in terms of the invariant form factors 
$F_i^{V/A}$,  where $\lambda_M = t, \pm 1, 0$ and 
$\lambda_2 = \pm 1/2, \pm 3/2$ are  the helicity components of the meson
$M\,(M=P,V)$ and the baryon $B_2$,  respectively. 
We need to calculate the expressions 
\be
H_{\lambda_2\lambda_M} = 
\la B_2(p_2,\lambda_2) |\bar q_2 O_\mu q_1 | B_1(p_1,\lambda_1) \ra
\epsilon^{\dagger\,\mu}(\lambda_M) 
=  H_{\lambda_2\lambda_M}^V - H_{\lambda_2\lambda_M}^A
\en 
where we split the  helicity amplitudes into their vector and axial parts.
For the color enhanced decays the operator  $\bar q_2 O_\mu q_1$ represents 
a charged current transition while, for the color suppressed decays, 
$\bar q_2 O_\mu q_1$ describes a neutral current transition.
We work in the rest frame of the baryon $B_1$ with the baryon $B_2$ 
moving in the positive $z$-direction:
$p_1 = (M_1, \vec{\bf 0})$, $p_2 = (E_2, 0, 0, |{\bf p}_2|)$ and 
$q = (q_0, 0, 0, - |{\bf p}_2|)$. The helicities of the three particles
are related by $\lambda_1 = \lambda_2 - \lambda_M$. We use the notation
$\lambda_P=\lambda_t=0$ for the scalar $(J=0)$ contribution in order to 
set the helicity label apart from $\lambda_V=0$ used for the longitudinal 
component of the $J=1$ vector meson. The relations connecting the helicity 
amplitudes to the invariant form factors are given 

\noindent
for the transition $\frac12^+ \to \frac12^+$\,: \quad
$H^V_{-\lambda_2,-\lambda_M} = + H^V_{\lambda_2,\lambda_M}$ and 
$H^A_{-\lambda_2,-\lambda_M} = - H^A_{\lambda_2,\lambda_M}$\,.

\bea 
\begin{array}{lcrlcl}
H_{\frac12 t}^V &=& \sqrt{Q_+/q^2} \, 
\Big( F_1^V M_- + F_3^V \frac{q^2}{M_1} \Big) \qquad 
&\qquad 
H_{\frac12 t}^A &=& \sqrt{Q_-/q^2} \, 
\Big( F_1^A M_+ - F_3^A \frac{q^2}{M_1} \Big) 
\\[1.1ex]
H_{\frac12 0}^V &=& \sqrt{Q_-/q^2} \,
\Big( F_1^V M_+ + F_2^V \frac{q^2}{M_1} \Big) \qquad 
& \qquad 
H_{\frac12 0}^A &=& \sqrt{Q_+/q^2} \,   
\Big( F_1^A M_- - F_2^A \frac{q^2}{M_1} \Big) 
\\[1.1ex]
H_{\frac12 1}^V &=& \sqrt{2Q_-} 
\Big( - F_1^V - F_2^V \frac{M_+}{M_1} \Big) \qquad 
& \qquad 
H_{\frac12 1}^A &=& \sqrt{2Q_+} \,  
\Big( - F_1^A + F_2^A \frac{M_-}{M_1} \Big) 
\\
\end{array}
\ena 

and for the transition $\frac12^+ \to \frac32^+$\,: \quad 
$H^V_{-\lambda_2,-\lambda_M} = - H^V_{\lambda_2,\lambda_M}$ and 
$H^A_{-\lambda_2,-\lambda_M} = +  H^A_{\lambda_2,\lambda_M}$. 

\bea 
H_{\frac12 t}^V &=& - \sqrt{\frac23\cdot \frac{Q_+}{q^2}} \, 
\frac{Q_-}{2M_1M_2} 
\Big( F_1^V M_1 - F_2^V M_+ + F_3^V \frac{M_+M_--q^2}{2M_1}  
+ F_4^V \frac{q^2}{M_1} \Big) 
\nn[1.1ex]
H_{\frac12 0}^V &=& - \sqrt{\frac23\cdot \frac{Q_-}{q^2}} \, 
\Big( F_1^V \frac{M_+M_--q^2}{2M_2} - F_2^V \frac{Q_+ M_-}{2M_1M_2}  
+ F_3^V \frac{|{\bf p_2}|^2}{M_2} \Big) 
\nn[1.1ex]
H_{\frac12 1}^V &=& \sqrt{\frac{Q_-}{3}} \,  
\Big( F_1^V - F_2^V \frac{Q_+}{M_1M_2} \Big) 
\qquad
H_{\frac32 1}^V = -  \, \sqrt{Q_-} \, F_1^V 
\\[2ex]
H_{\frac12 t}^A &=& \sqrt{\frac23\cdot \frac{Q_-}{q^2} }
\frac{Q_+}{2M_1M_2} 
 \Big( F_1^A M_1 + F_2^A M_- + F_3^A \frac{M_+M_--q^2}{2M_1}   
+ F_4^A \frac{q^2}{M_1}\Big) 
\nn[1.1ex]
H_{\frac12 0}^A &=&  \sqrt{\frac23\cdot \frac{Q_+}{q^2} } 
\Big( F_1^A \frac{M_+M_--q^2}{2M_2} + F_2^A \frac{Q_-M_+}{2M_1M_2} 
+ F_3^A  \frac{|{\bf p_2}|^2}{M_2}  \Big) 
\nn[1.1ex]
H_{\frac12 1}^A &=& \sqrt{\frac{Q_+}{3}} 
\Big( F_1^A - F_2^A \frac{Q_-}{M_1M_2} \Big) 
\qquad
H_{\frac{3}{2}1}^A = \, \sqrt{Q_+} F_1^A 
\nonumber
\ena 
\noindent We use the abbreviations
$M_\pm = M_1 \pm M_2$, 
$Q_\pm = M_\pm^2 - q^2$. The magnitude of the momentum of the daughter
baryon $B_2$ is given by
${|\bf p_2|} = \sqrt{Q_+Q_-}/2M_1= \lambda^{1/2}(M_1^2,M_2^2,q^2)/(2M_1)$. 
\vspace{0.2cm}

Let us add a few remarks on the helicity composition of the vector and axial
vector helicity amplitudes. At the zero recoil point $q^2=(M_1-M_2)^2$
the vector helicity amplitudes vanish and the transverse-to-longitudinal
composition can be seen to be given by
${\cal F}_L/{\cal F}_T=1/2$ for both $1/2^+\to 1/2^+,3/2^+$ transitions
(``allowed Fermi-Teller transition'').
At the other end of the $q^2$-spectrum at $q^2=0$ the longitudinal
mode dominates. These findings have a bearing on the transverse-to-longitudinal
composition of the vector mesons in the nonleptonic decays to be discussed
later on.

Using the helicity amplitudes one can write down very compact expressions
for the various decay rates.
The semileptonic decay width is given by ($m_\ell=0$)
\be
\label{semilep}
\Gamma(B_1 \to B_2\,+\,\ell^+ \,\nu_\ell)=\int_0^{(M_1-M_2)^2} dq^2 \,\,
\frac{d\,\Gamma(B_1 \to B_2\,+\,\ell^+ \,\nu_\ell)}
{dq^2}
\en
where
\be
\label{semilep2}
\frac{d\,\Gamma(B_1 \to B_2\,+\,\ell^+ \,\nu_\ell)}{dq^2} 
= \frac{1}{192\pi}G_F^2 \, \frac{|{\bf p_2}|q^2}{M_1^2} \, 
|V_{ij}|^2 \,{\cal H}_V \, ({\cal H}'_V) \,.
\en
For the nonleptonic decays one has
 \be
\Gamma(B_1 \to B_2\,+\,V) 
= \frac{G_F^2}{32 \pi} \, \frac{|{\bf p_2}|}{M_1^2} \, 
|V_{ij} V^\ast_{kl}|^2 \, C_{\rm eff}^2 \, f_V^2 \, M_V^2 
\,{\cal H}_V\, ({\cal H}'_V)\,,
\label{nldecay1}
\en
\be
\Gamma(B_1 \to B_2\,+\,P) 
= \frac{G_F^2}{32 \pi} \, \frac{|{\bf p_2}|}{M_1^2} \, 
|V_{ij} V^\ast_{kl}|^2 \, C_{\rm eff}^2 \, f_P^2 \, M_P^2 
\,{\cal H}_S \,({\cal H}'_S)\,,  
\label{nldecay2}
\en
where we denote the sum of the squared moduli of the  helicity amplitudes
by ${\cal H}_V$, ${\cal H}_S$, ${\cal H}'_V$ and ${\cal H}'_S$ according to
the two cases

\bea
\label{trace}
1/2^+ &\to& 1/2^+\,: \qquad {\cal H}_V= 
\sum_{\lambda_2=\pm 1/2,\,\lambda_V=\pm1,0}|H_{\lambda_2,\lambda_V}|^2  
\qquad
   {\cal H}_S= \sum_{\lambda_2=\pm1/2}|H_{\lambda_2,\lambda_t}|^2 \nn
1/2^+ &\to& 3/2^+\,: \qquad {\cal H}'_V= 
\sum_{\lambda_2=\pm 1/2,\pm3/2,\,\lambda_V=\pm1,\,0}
|H_{\lambda_2,\lambda_V}|^2 \qquad 
{\cal H}'_S= \sum_{\lambda_2=\pm1/2}|H_{\lambda_2,\lambda_t}|^2 
\ena
Angular momentum conservation dictates the 
constraint $|\lambda_2 -\lambda_M|\le 1/2$ for the helicity amplitudes 
since the initial state baryon has spin $1/2$.

It is quite convenient to work with normalized helicity amplitudes which 
we denote by $\hat H_{\lambda_2\,\lambda_M}$. The helicity amplitudes 
are normalized according to 
\be\label{norm}
\hat H_{\lambda_2\,t} = H_{\lambda_2\,t}\,/\,{\cal H}_S^{1/2}
\qquad \qquad \hat H_{\lambda_2\,\lambda_V} = 
H_{\lambda_2\,\lambda_V}\,/\,{\cal H}_V^{1/2} 
\en
for the $1/2^+ \to 1/2^+$ case and accordingly for the $1/2^+ \to 3/2^+$ case
with ${\cal H}_S \to {\cal H}'_S$ and ${\cal H}_V \to {\cal H}'_V$.

\section{Interpolating currents and calculation of the transition 
form factors in the confined covariant quark model (CCQM)}

As described in the introduction we use the confined covariant quark model 
(CCQM) to calculate the various $1/2^+ \to 1/2^+$ and $1/2^+ \to 3/2^+$ 
transition form factors $F^{V,A}_i(q^2)$ that are needed in the calculation 
of the helicity amplitudes. We describe the coupling of the baryons with the 
constituent quarks by nonlocal extensions of the interpolating currents 
(see details in Refs.~\cite{Faessler:2001mr,Gutsche:2017hux}, 
\cite{Ivanov:1997ra}-\cite{Ivanov:1996fj}).    
In Table~\ref{intpoltable} we list the
interpolating currents needed in the present application.
   
\begin{table}[ht]
    \caption
    {Cabibbo-favored nonleptonic two-body decays of double
      heavy charm baryons including $W$--exchange contributions.}
   \vspace{0.2cm}
\begin{center}
\begin{tabular}{l|c|c|c}
  \hline
  \hline
Baryon\qquad &\quad $J^P$ \quad &Interpolating current & Mass [MeV] \\ [1.2ex]
      \hline
      $\Xi_{cc}^{++}$\quad \quad \quad
      & $\frac 12^+$ & $\epsilon^{abc}\gamma^\mu\gamma_5 \ u^a c^b
   C\gamma_\mu c^c$  & 3620.6 
      \\[0.5ex]
      $\Xi_{cc}^{+}$\quad \quad \quad& $\frac 12^+$
      &$\epsilon^{abc}\gamma^\mu\gamma_5 \ d^a c^b
   C\gamma_\mu c^c$ &$3620.6$ 
      \\[0.5ex]
      $\Omega_{cc}^{+}$\quad \quad \quad & $\frac 12^+$ &
      $\epsilon^{abc}\gamma^\mu\gamma_5 \ s^a c^b
   C\gamma_\mu c^c$ & 3710
      \\[0.5ex]
$\Sigma_c^{++}$  \quad & $\frac 12^+$ 
& $\epsilon^{abc}\gamma^\mu\gamma_5 \ c^a u^bC\gamma_\mu u^c$ & 2453.97 
      \\[0.5ex]
$\Sigma_c^{*++}$ & $\frac 32^+$ & $\epsilon^{abc} 
\ c^a u^bC\gamma_\mu u^c $ & 2518.41
\\[0.5ex]
$\Xi_{c}^{+}$ & $\frac 12^+$ & $\epsilon^{abc} c^a
u^bC\gamma_5 s^c$ & 2467.93 
\\[0.5ex]
$\Xi_{c}^{'+}$ & $\frac 12^+$ & $\epsilon^{abc} \gamma^\mu\gamma_5 c^a
u^bC\gamma_\mu s^c$ & 2577.4
\\[0.5ex]
$\Xi_{c}^{0}$ & $\frac 12^+$ & $\epsilon^{abc} c^a
d^bC\gamma_5 s^c$ & 2470.85
\\[0.5ex]
$\Xi_{c}^{'0}$ & $\frac 12^+$ & $\epsilon^{abc} \gamma^\mu\gamma_5 c^a
d^bC\gamma_\mu s^c$ & 2577.9
\\[0.5ex]
$\Xi_{c}^{*+}$ & $\frac 32^+$ & $\epsilon^{abc} c^au^bC\gamma_\mu s^c$ 
& 2645.57
 \\[0.5ex]
$\Xi_{c}^{*0}$ & $\frac 32^+$ & $\epsilon^{abc} c^ad^bC\gamma_\mu s^c$ 
& 2646.38
 \\[0.5ex]
$\Omega_c^0$ & $\frac 12^+$ & $ \epsilon^{abc} 
\gamma^\mu\gamma_5 c^a s^b C \gamma_\mu s^c$ & 2695.2
 \\[0.5ex]
 $\Omega_c^{*0}$ & $\frac 32^+$ &$\epsilon^{abc} c^a s^b 
C \gamma_\mu s^c$ & 2765.9
 \\[0.5ex]
\hline
\hline
\end{tabular} 
\end{center}
\label{intpoltable}
  \end{table}
The three constituent 
quarks are treated as separate dynamic entities which propagate with  
fully covariant fermion propagators $S_q(k)=1/(m_q-k\!\!\! /)$ in the
two-loop Feynman diagram which describes the current-induced transition
between the respective baryons. The propagator masses $m_q$ are
constituent quark masses fixed 
in previous analyses of a multitude of hadronic processes within our 
approach (see, e.g., Refs.~\cite{Gutsche:2013oea,Gutsche:2017wag}). 

Apart from the choice of the interpolating current and the constituent  
quark masses there are two parameters that describe the structure of 
a baryon in the CCQM. These are the coupling factor of the baryon to 
its constituent quarks $g_B$ and the size parameters $\Lambda_B$  
characterizing the size of the nonlocal distribution of the quarks in the 
baryons. The coupling factor $g_B$ and the size parameter $\Lambda_B$ become 
related by the compositeness condition of Salam and 
Weinberg~\cite{Salam:1962ap,Weinberg:1962hj}. By analogy we treat mesons 
as bound states of a constituent quark and an antiquark, i.e. we construct 
respective nonlocal interaction Lagrangians of mesons with their constituent 
quarks (see details in Refs.~\cite{Branz:2009cd,Dubnicka:2013vm}). 

The details of calculating the $1/2^+ \to 1/2^+$ and $1/2^+ \to 3/2^+$ 
transition form factors between baryons have been discussed in detail 
in Refs.~\cite{Faessler:2001mr,Gutsche:2017hux}, 
\cite{Ivanov:1997ra}-\cite{Ivanov:1996fj} and need not be repeated here. 

\section{Polarization, longitudinal/transverse helicity fractions 
and angular decay distributions}

Since the semileptonic and nonleptonic two-body decays of the 
$\Xi_{cc}^{++},\,\Xi_{cc}^{+}$ 
and $\Omega_{cc}^+$ are mediated by weak interactions one can expect 
sizable polarization effects in these decays entailing nontrivial  
angular decay distributions in the decays of the mesons and baryons  
further down the decay chains. We treat the initial state 
baryons $\Xi_{cc}^{++},\,\Xi_{cc}^{+}$ and $\Omega_{cc}^+$ as being 
unpolarized. In principle, the
parent baryons could acquire a nonzero transverse polarization in the hadronic
production process which would depend on the rapidity of the baryon in
question. However, since one is
usually averaging over the rapidities
of the production process, the parent baryons become effectively unpolarized
(for more details see~\cite{Gutsche:2017wag}).

\subsection{Semileptonic decays}

We only consider the Cabibbo-favored semileptonic decays of the double 
heavy charm baryons $\Xi_{cc}^{++},\,\Xi_{cc}^{+}$ and $\Omega_{cc}^+$  
induced by the quark level $c \to s$ transition.  
The Cabibbo-suppressed semileptonic decays induced by the quark level  
$c \to d$ transitions are suppressed by an overall factor 
$(V_{cd}/V_{cs})^2=0.049$. The Cabibbo suppression factor $(V_{cd}/V_{cs})^2$ 
is partly offset by the larger phase space of the $\Delta S=1$ 
Cabibbo-suppressed decays which then amounts to an overall suppression 
factor of $\sim 0.1$ (see e.g. \cite{Wang:2017mqp}).

The Q values of the semileptonic $c \to s$ decays discussed here are not 
large enough to allow for the semileptonic $\tau$ modes. On the other hand, 
the Q values are sufficiently large to allow one to neglect the lepton masses 
in the semileptonic $e^+$- and $\mu^+$-modes. 

\begin{table}[ht]
   \caption{\label{tablesl}
Cabibbo-favored semileptonic decays of double heavy 
charm baryons induced by the charm level $c \to s$ transition
   ($\ell=e^+,\,\mu^+$).}
   \vspace{0.2cm}
\begin{center}
\begin{tabular}{llcccccc}
  \hline
  \hline
\phantom{$1/2^+ \to 1/2^+$}&\phantom{$\hat{\hat{I}_3}$} &
$ \Gamma$ [$10^{-13}$ GeV] &
$\quad\cal{B}$ $[\%]$ &
$\qquad \langle{\cal F}_+\rangle$&\quad$\langle{\cal F}_0\rangle$&
\quad$\langle{\cal F}_-\rangle$&
\phantom{$\alpha_P^{(n)}(O(\alpha^2_s))$} \\ 
      \hline
   $1/2^+ \to 1/2^+$ \quad \quad& 
$\Xi_{cc}^{++} \to \Xi_c^+ + \ell^+ \nu_\ell $ 
& 0.70 & 2.72 & 0.02 & 0.88 & 0.10 
\\[0.5ex]
\phantom{$1/2^+ \to 1/2^+$}&$\Xi_{cc}^{++} \to 
\Xi_{c}^{'+} + \ell^+ \nu_\ell $ 
& 0.97 & 3.76 & 0.09 & 0.55 & 0.36
 \\[0.5ex]
\phantom{$1/2^+ \to 1/2^+$}&$\Xi_{cc}^{+} \to \Xi_{c}^{0} + \ell^+ 
\nu_\ell $ 
& 0.69 & 2.00 & 0.02 & 0.88 & 0.10 
\\[0.5ex]
\phantom{$1/2^+ \to 1/2^+$}&$\Xi_{cc}^{+} \to \Xi_{c}^{'0} + \ell^+  
\nu_\ell $ 
& 0.97 & 2.79 & 0.09 & 0.55 & 0.36
\\[0.5ex]
\phantom{$1/2^+ \to 1/2^+$}&
$\Omega_{cc}^{+} \to \Omega_{c}^{0} + \ell^+  \nu_\ell $ 
& 1.82 & 7.07 & 0.09 & 0.55 & 0.36 
\\[0.5ex]
$1/2^+ \to 3/2^+$ &$\Xi_{cc}^{++} \to \Xi_{c}^{*+} + \ell^+ 
\nu_\ell $ 
& 0.22 & 0.86 & 0.12 & 0.49 & 0.39 
\\[0.5ex]
\phantom{$1/2^+ \to 3/2^+$} &$\Xi_{cc}^{+} \to \Xi_{c}^{*0} + \ell^+ 
\nu_\ell $ 
& 0.22 & 0.64 & 0.12 & 0.49 & 0.39 
\\[0.5ex]
\phantom{$1/2^+ \to 3/2^+$}&$\Omega_{cc}^{+} \to \Omega_{c}^{*0} + \ell^+ 
\nu_\ell $ 
& 0.40 & 1.27 & 0.12 & 0.49 & 0.39
\\[0.5ex]
\hline
\hline
\end{tabular} 
\end{center}
\end{table}
   
In Table~\ref{tablesl} we present our numerical results for the  
Cabibbo-favored semileptonic decays of the double heavy charm baryon  
states $\Xi_{cc}^{++},\,\Xi_{cc}^{+}$ and $\Omega_{cc}^+$. We also list 
branching fractions for the semileptonic decays of the $\Xi_{cc}^{++}$ 
based on the recent measurement of the lifetime of the 
$\Xi_{cc}^{++}$~\cite{Aaij:2018wzf}: 
$\tau_{\Xi_{cc}^{++}}= (256^{+24}_{-22} \pm 14)$ fs. 

For the semileptonic decays of the $\Xi_{cc}^{+}$ and $\Omega_{cc}^+$ 
we quote nominal branching fractions. These are nominal since the 
lifetimes of the double heavy charm baryon states $\Xi_{cc}^{+}$ and 
$\Omega_{cc}^+$ have not been measured yet. One has to rely on theoretical 
calculations~\cite{Kiselev:1998sy, Chang:2007xa} from which we take the 
median values 
\be
\tau_{\Xi_{cc}^{+}}= 190\,{\rm fs}\,, \qquad
\tau_{\Omega_{cc}^{+}}= 210\,{\rm fs}\,. 
\en
In the case that the experimental lifetime of the $\Xi_{cc}^{++}$ changes
in the future and the lifetimes of the $\Xi_{cc}^{+}$ and $\Omega_{cc}^+$ 
become known one has to rescale our branching fractions by the ratios
\be
\left(\frac{\tau_{\Xi_{cc}^{++}}}{256\,{\rm fs}}\right)\,, \qquad
\left(\frac{\tau_{\Xi_{cc}^{+}}}{190\, {\rm fs}}\right)\,,  \qquad
\left(\frac{\tau_{\Xi_{cc}^{++}}}{210\,{\rm fs}}\right)\,. 
\en
In Table~\ref{tablesl} we also include numerical values for the 
$q^2$--averages of the 
transverse-plus, longitudinal and transverse-minus helicity fractions of 
the off-shell gauge boson $W^-$ where we denote the averages by 
$\langle{\cal F}_+\rangle,\,\langle{\cal F}_0\rangle$ 
and $\langle{\cal F}_-\rangle$. When taking the $q^2$--averages one has to 
integrate the numerators and denominators separately including the factor 
$|{\bf p_2}|q^2$ (see Eq.(\ref{semilep2})). The $q^2$--dependent helicity  
fractions are defined by
\bea
\label{trace2}
1/2^+ &\to& 1/2^+\,: \qquad {\cal F}_+(q^2)= |\hat H_{1/2,1}|^2  \qquad
{\cal F}_0(q^2)= |\hat H_{1/2,0}|^2+|\hat H_{-1/2,0}|^2  \qquad 
{\cal F}_-(q^2) = |\hat H_{-1/2,-1}|^2 \nn
1/2^+ &\to& 3/2^+\,:
\qquad {\cal F}_+(q^2)= \sum_{\lambda_2} |\hat H_{\lambda_2,1}|^2  \qquad
       {\cal F}_0(q^2)= |\hat H_{1/2,0}|^2+|\hat H_{-1/2,0}|^2
       \qquad {\cal F}_-(q^2)= \!\!\sum_{\lambda_2} |\hat H_{\lambda_2,-1}|^2 
\ena
where $\lambda_2=1/2,3/2$ for ${\cal F}_+(q^2)$ and 
$\lambda_2=-1/2,-3/2$ for ${\cal F}_-(q^2)$ in the $1/2^+ \to 3/2^+$ case.  
Since we use  normalized helicity amplitudes $\hat H_{\lambda_2\,\lambda_M}$ 
[see Eq.(\ref{norm})] the helicity fractions satisfy 
${\cal F}_+(q^2)+{\cal F}_0(q^2)+{\cal F}_-(q^2)=1$.  
The angular decay distribution of the lepton $\ell^-$ in the  
$(\ell^-;\bar \nu_\ell)$ rest frame is given by
\be
W(\theta )= \frac 38 (1-\cos\theta)^2 {\cal F}_+ +
\frac 34 \sin^2\theta \,{\cal F}_0 +\frac 38 (1+\cos\theta)^2 {\cal F}_- 
\en
where the angle $\theta$ is defined in analogy to the angle $\theta_V$ in 
Fig.~\ref{fig:Cascade_NLD} with the change of labeling $(K^- \to \ell^-)$,   
$(\pi^+ \to \bar \nu_\ell)$ and $(\bar K^{*0} \to W^-_{\rm off-shell})$.  
We do not discuss polarization effects on the hadron side of the 
semileptonic decays. These can be  
discussed  along the lines of~\cite{Gutsche:2015rrt}. 

\subsection{Nonleptonic decays}

We discuss the rates, branching fractions and angular decay distributions 
of the four classes of decays 
\bea
   1/2^+ &\to& 1/2^+ + 0^- \nn
   1/2^+ &\to& 1/2^+ + 1^- \nn
   1/2^+ &\to& 3/2^+ + 0^- \nn
   1/2^+ &\to& 3/2^+ + 1^- \,\,.
\ena
Each of the above classes contains four factorizing nonleptonic two-body 
decays. Thus we iscuss 
altogether 16 factorizing nonleptonic two-body decays comprising the decays  
$\Xi_{cc}^{++} \to \Sigma_{c}^{(*)++} + \bar K^{(*)0 }$,  
$\Omega_{cc}^{+} \to \Omega_c^{(*)0} + \pi^+(\rho^+)$, 
$\Xi_{cc}^{++} \to \Xi_{c}^{\prime(*)+} + \pi^+(\rho^+)$  
and $\Omega_{cc}^{+} \to \Xi_{c}^{\prime (*)+} +\bar {K}^{(*)0}$ as they
appear in rows 1, 2, 13 and 15 of~Table \ref{topol}.

When discussing angular decay distributions we concentrate 
on the $1/2^+ \to 1/2^+ + 1^-$ two-sided cascade decay 
$\Xi_{cc}^{++} \to \Sigma_c^{++}(2455;1/2^+) (\to \Lambda_c^+ \pi^+) 
+ \bar K^{*0} (\to K^-  \pi^+)$ and the $1/2^+ \to 3/2^+ + 1^-$ 
cascade decay 
$\Xi_{cc}^{++} \to \Sigma_c^{++}(2520;3/2^+) (\to \Lambda_c^+ \pi^+) 
+ \bar K^{*0} (\to K^-  \pi^+)$ as well as the corresponding two one-sided 
cascade decays with $\bar K^{*0}$ replaced by $\bar K^{0}$. 
These decay chains are favored from an experimental point of view 
since the second stage branching ratios are large. On the baryon side  
the daughter baryon decays $\Sigma_c^{++}(2455,1/2^+) \to \Lambda_c^+ \pi^+$ 
and $\Sigma_c^{++}(2520;3/2^+) \to \Lambda_c^+ \pi^+$  have a large branching 
ratio close to $ 100\,\%$. On the meson side the branching ratio of the decay  
$\bar K^{*0} \to K^-  \pi^+$ is also quite large ($\sim 66\,\%$ from isospin 
invariance). Further, all final states in the decay chains are charged which 
is optimal from an experimental point of view. 

In the following we discuss the classes of decays separately: 
\begin{itemize}
\item \quad $1/2^+ \to 1/2^+ + 0^-$ 

In Table~\ref{tablenum1} we list the rates, branching fractions and the
polarization of the daughter baryon $P_{B_2}$ for the four decays in this  
class. 

\begin{table}[ht]
\caption
{\label{tablenum1}Cabibbo favored factorizing nonleptonic two-body 
decays of double heavy charm baryons 
induced by the quark level $c \to s;d \to u$ transitions for the
cases $1/2^+ \to 1/2^+ +0^-$.}
   \vspace{0.2cm}
\begin{center}
\begin{tabular}{llccc}
  \hline
  \hline
\phantom{$1/2^+ \to 1/2^+ +0^-$}&\phantom{$\hat{\hat{I}_3}$} 
& \,\,\, $\Gamma$ [$10^{-13}$ GeV] \,\,\, 
& \,\,\, $\cal{B}$ [$\%$] \,\,\, 
& \,\,\, $P_{B_2}$ \,\,\, \\ 
      \hline
$1/2^+ \to 1/2^+ +0^-$ \quad \quad& $\Xi_{cc}^{++} \to \Sigma_c^{++}
      +\bar K^0$ & 0.32 & 1.25 & $-0.96$ 
\\[0.5ex]
      \phantom{$1/2^+ \to 1/2^+ + 0^-$}&$\Xi_{cc}^{++} \to \Xi_{c}^{'\,+} +
 \pi^+$ & 0.78 & 3.03 & $-0.94$ 
 \\[0.5ex]
\phantom{$1/2^+ \to 1/2^+ + 0^-$}&$\Omega_{cc}^{+} \to  
\Xi_{c}^{'\,+} +\bar K^0$ & 0.17  & 0.54 & $-0.97$ 
 \\[0.5ex]
 \phantom{$1/2^+ \to 1/2^+ + 0^-$}&$\Omega_{cc}^{+} \to \Omega_{c}^{0} +
 +\pi^+$ & 1.58 & 5.05 & $-0.94$ 
 \\[0.5ex]
\hline
\hline
\end{tabular} 
\end{center}
 \end{table}

As mentioned before we concentrate on the cascade decay 
$\Xi_{cc}^{++} \to \Sigma_c^{++}(\to \Lambda_c^+ \pi^+)+ \bar K^{0}$ 
when discussing polarization effects and angular decay distributions. 
The stage 2 decay $\Sigma_c^{++}\to \Lambda_c^+ \pi^+$ is a parity-conserving 
strong decay such that the one-fold angular decay distribution of this 
cascade decay is given by
\bea
\label{angdist1}
W(\theta_B)&=& \sum_{\lambda_2,\,\lambda_3=\pm1/2}
\hat H_{\lambda_2\,t}\hat H^*_{\lambda_2\,t}
d^{1/2}_{\lambda_2\,\lambda_3}(\theta_B)
d^{1/2}_{\lambda_2\,\lambda_3}(\theta_B) \nn
&=&\sum_{\lambda_2=\pm1/2}
\hat H_{\lambda_2\,t}\hat H^*_{\lambda_2\,t}=1
\ena
where the polar angle $\theta_B$ is defined in Fig.~\ref{fig:Cascade_NLD}.

\begin{figure}[htb]
\begin{center}
\epsfig{figure=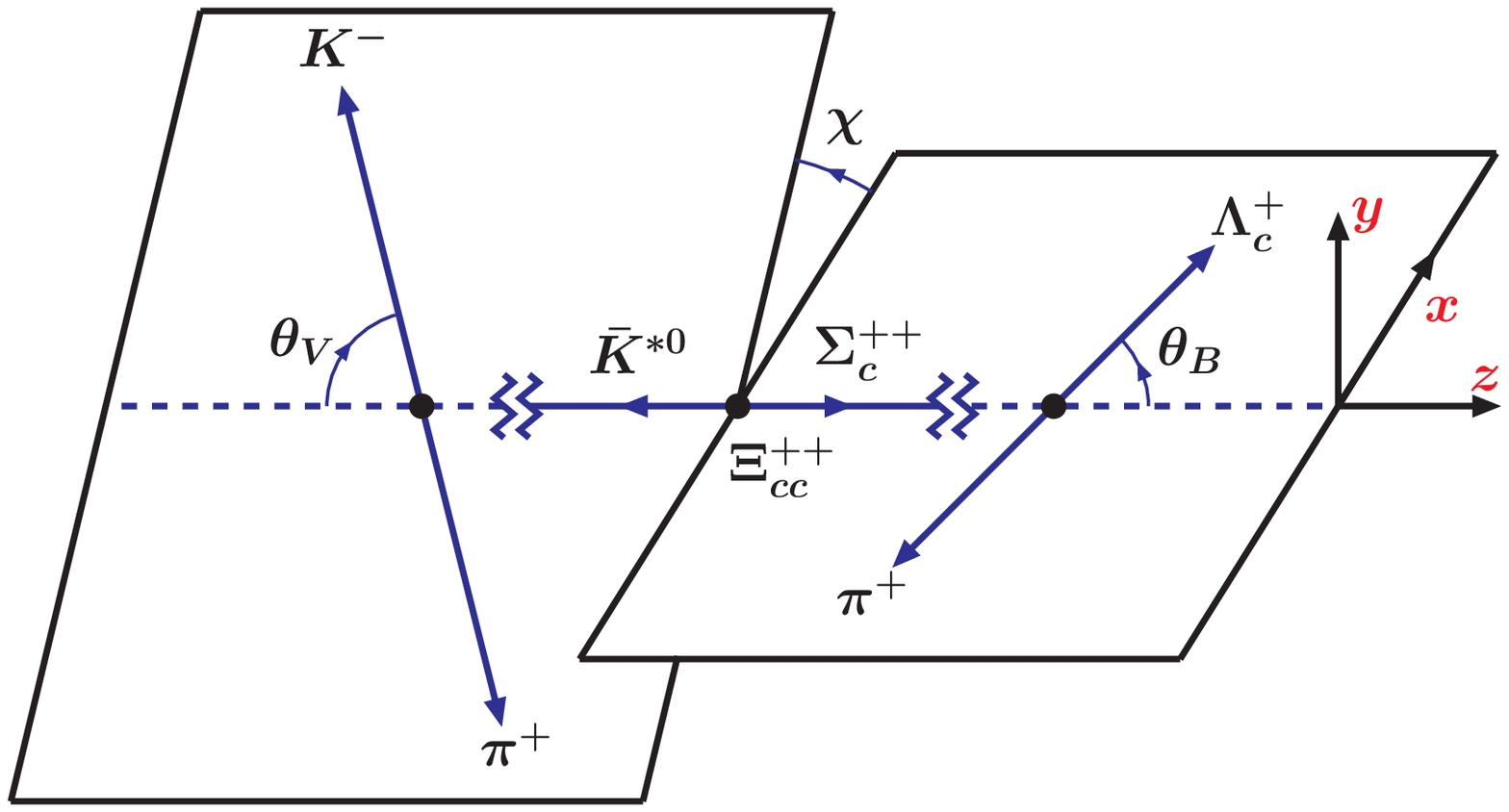,scale=.8}
\caption{Definition of the angles $\theta_B$, $\theta_V$,
and $\chi$ in the cascade decay
$\Xi_{cc}^{++} \to \Sigma_c^{++}(\to \Lambda_c^+ \pi^+)                     
+ \overline{K}^{0*}(\to K^- \pi^+)$. The $\Sigma_c^{++}$ can either be the
$\Sigma_c^{++}(2455;1/2^+)$ or the $\Sigma_c^{++}(2520;3/2^+)$. 
\label{fig:Cascade_NLD}}
\end{center}
\end{figure}

When simplifying the decay distribution~(\ref{angdist1}) 
we have used the orthonormality relation for the spin 1/2 Wigner
$d^{1/2}$--function
\be
\label{orthonorm}
\sum_{\lambda_3} d^{1/2}_{\lambda_2\,\lambda_3}(\theta_B)
d^{1/2}_{\lambda'_2\,\lambda_3}(\theta_B)
=\delta_{\lambda_2\,\lambda'_2} \,.
\en
 The bilinear forms of the helicity
amplitudes sum up to 1 since normalized helicity amplitudes are used. 
The angular decay distribution~(\ref{angdist1}) can be seen to be flat.

The longitudinal polarization of the daughter
baryon $\Sigma_c^{++}$ in the stage 1 decay
$\Xi_{cc}^{++} \to \Sigma_c^{++} + \bar K^{0}$ is given by
\be
   P_{\Sigma_c^{++}}=
       |\hat H_{\frac12 t}|^2 -|\hat H_{-\frac12 t}|^2\,. 
\en
We will refer to this polarization parameter as $P_{B_2}$ in the general 
context and list $P_{B_2}$ for the other three decays in 
Table~\ref{tablenum1}. As discussed after~(\ref{angdist1}) the polarization 
of the daughter baryon $P_{\Sigma_c^{++}}$ cannot be resolved from its 
angular  decay distribution because the decay 
$\Sigma_c^{++}\to \Lambda_c^+ \pi^+$ is a strong decay. 
However, the polarization of the $\Sigma_c^{++}$ is transferred to the 
second stage baryon $\Lambda_c^+$. The degree of polarization transfer 
depends on the baryon side polar angle $\theta_B$ and is given by
\be
\label{poltransf}
   P_{\Lambda_c^{+}}=P_{\Sigma_c^{++}} \cos\theta_B \,.
\en
The polarization of the $\Lambda_c^+$ can in turn be analyzed through its 
weak decays as e.g. in the decay $\Lambda_c^+ \to \Lambda^0\,\pi^+$ which 
possesses a large analyzing power of $-0.91 \pm 0.15$~\cite{Tanabashi:2018oca}. 
From an experimental point of view the decay 
$\Lambda^+_c \to p K^- \pi^+$ would be preferred as an analyzing channel 
since it has a larger branching fraction than the decay 
$\Lambda_c^+ \to \Lambda^0\,\pi^+$ by a factor of $\sim \!5$. 
However, to our knowledge the analyzing power of this mode has neither been 
measured experimentally nor calculated theoretically except for an analysis 
of the two subchannels $\Lambda_c^+ \to p \bar K^{*0}$ and  
$\Lambda_c^+ \to \Delta^{++}K^-$~\cite{Konig:1993wz}.

The decays $\Xi_{cc}^{++} \to \Xi_{c}^{'\,+} +\pi^+$ and  
$\Omega_{cc}^{+} \to  \Xi_{c}^{'\,+} +\bar K^0$ involve daughter charm 
baryon state $\Xi_{c}^{'\,+}$ which then cascades down to the ground state 
$\Xi_{c}^{+}$ via a parity-conserving one-photon emission 
$\Xi_{c}^{'\,+} \to \Xi_{c}^{+} +\gamma$. As in the decay 
$\Xi_{cc}^{++} \to \Sigma_c^{++}(\to \Lambda_c^+ \pi^+)+ \overline{K}^{0}$ 
discussed above the helicity angle distribution of the $\Xi_{c}^{+}$ is 
flat. Differing from~Eq.~(\ref{poltransf}) the polarization transfer is now 
\be
   \label{reversal}
   P_{\Xi_{c}^{+}}=-P_{\Xi_{c}^{'\,+}} \cos\theta_B\, .
\en
As concerns the decay $\Omega_{cc}^{+} \to \Omega_{c}^{0} +\pi^+$ 
the daughter baryon $\Omega_{c}^{0}$  
has a multitude of decay channels of which the relevant decay asymmetries 
have not been determined yet experimentally. Theoretical predictions for 
the two-body decay asymmetries of the the daughter baryon 
$\Omega_{c}^{0}$  can be found in Refs.~\cite{Korner:1978tc,Korner:1992wi}.

\item \quad $1/2^+ \to 1/2^+ + 1^-$ \\[10pt]
In Table~\ref{tablenum2} we list the rates, branching fractions, the  
polarization of the daughter baryon $P_{B_2}$ and the helicity fractions  
of the vector meson for the four decays in this class.

\begin{table}[ht]
\caption
    {\label{tablenum2}Cabibbo-favored factorizing nonleptonic two-body 
decays of double heavy charm baryons induced by the quark level 
$c \to s;d \to u$ transitions for the case
      $1/2^+ \to 1/2^+ + 1^-$.}
   \vspace{0.2cm}
\begin{center}
\begin{tabular}{llccccc}
  \hline
  \hline
\phantom{$1/2^+ \to 1/2^+ +0^-$}&\phantom{$\hat{\hat{I}_3}$} 
& \,\,\, $\Gamma$ [$10^{-13}$ GeV] \,\,\, 
& \,\,\, $\cal{B}$ [$\%$] \,\,\, 
& \,\,\, ${\cal F}_L$ \,\,\, 
& \,\,\, ${\cal F}_T$ \,\,\, 
& \,\,\, $P_{B_2}$ \,\,\, \\ 
      \hline
$1/2^+ \to 1/2^+ + 1^-$&$\Xi_{cc}^{++} \to \Sigma_c^{++}
 +\bar K^{*0}$ & 1.44 & 5.61 & 0.47 & 0.53 & $-0.82$ 
\\[0.5ex]
 \phantom{$1/2^+ \to 1/2^+ + 0^-$}&$\Xi_{cc}^{++} \to \Xi_{c}^{\prime\,+} +
 \rho^+$ & 4.14 & 16.10 & 0.49 & 0.51 & $-0.74$ 
 \\[0.5ex]
 \phantom{$1/2^+ \to 1/2^+ +1^-$}&
$\Omega_{cc}^{+} \to \Xi^{\prime +}_{c}+\bar K^{*0} $
 & 0.75 & 2.39 & 0.45 & 0.55 & $-0.79$ 
\\[0.5ex]  
\phantom{$1/2^+ \to 1/2^+ +1^-$}&$\Omega_{cc}^{+} \to \Omega_{c}^{0} +
 \rho^+$ & 8.29 & 26.44 & 0.48 & 0.52 & $-0.71$ 
\\[0.5ex]  
\hline
\hline
\end{tabular} 
\end{center}
\end{table}

The threefold angular decay distribution for the generic cascade decay 
$1/2^+ \to 1/2^+ (\to 1/2^+ +0^-)+ 1^-(\to 0^-+0^-)$ is given by
     \bea
\label{angdist2}
W(\theta_B, \theta_V, \chi)&=& \sum_{\lambda_V,\lambda'_V\lambda_2,\,
\lambda'_2,\,\lambda_3}
\delta_{\lambda_2-\lambda_V,\,\lambda'_2-\lambda'_V}
e^{-i(\lambda_V-\lambda'_V)\chi}
d^1_{0,\,\lambda_V}(\theta_V)d^1_{0\,\lambda'_V}(\theta_V)
\hat H_{\lambda_2\,\lambda_V}\hat H^*_{\lambda'_2\,\lambda'_V}
d^{1/2}_{\lambda_2\,\lambda_3}(\theta_B)
d^{1/2}_{\lambda'_2\,\lambda_3}(\theta_B)\nn
&=&\sum_{\lambda_V,\lambda_2}
d^1_{0,\,\lambda_V}(\theta_V)d^1_{0\,\lambda_V}(\theta_V)
\hat H_{\lambda_2\,\lambda_V}\hat H^*_{\lambda_2\,\lambda_V}\,.
\ena
where we have assumed that the stage 2 decays on the baryon and meson side
are strong and thus parity conserving as in the cascade decay
$\Sigma_c^{++}(2455;1/2^+)
(\to \Lambda_c^+ \pi^+)+ \bar K^{*0} (\to K^-  \pi^+)$ which we focus on
in the following. When evaluating the helicity sum in~(\ref{angdist2}) one
has to keep in mind that $|\lambda_2 -\lambda_V|\le 1/2$. The three angles
$\theta_V$, $\theta_B$ and $\chi$ describing the angular structure of the decay
are defined in Fig.~\ref{fig:Cascade_NLD}. We have again
used the orthonormality property Eq.~(\ref{orthonorm}) in the reduction of
the first row of Eq.~(\ref{angdist2}). The seeming threefold
angular decay distribution has collapsed to a onefold angular decay
distribution. In particular, the  azimuthal correlation between
the two decay planes spanned by $ \{\Lambda_c^+,\pi^+\}$ and $\{K^-, \pi^+\}$
vanishes.

The vector mesons on the meson side of the decay chain can be
transversely and
longitudinally polarized. We define the corresponding helicity fractions by 
\be
   {\cal F}_L = |\hat H_{\frac 12 0}|^2+|\hat H_{-\frac 12 0}|^2
    \qquad
   {\cal F}_T = |\hat H_{\frac 12 1}|^2 +|\hat H_{ -\frac 12 -1}|^2 \,. 
\en
For the meson-side decay $1^- \to 0^- +0^-$ one obtains the angular
decay distribution
\be
\label{angdis1}
W(\theta_V) =
\bigg(\frac 32 \cos^2\theta_V \,{\cal F}_L 
+\frac 34 \sin^2\theta_V \,{\cal F}_T \bigg) \,. 
\en
The longitudinal polarization of the daughter baryon $\Sigma_c^{++}$ 
depends on the the polar emission angle $\theta_V$ via 
\be
   \label{polsigma}
P_{B_2}(\cos\theta_V)=   P_{\Sigma_c^{++}}(\cos\theta_V)=\frac{
    \frac 34 \sin^2\theta_V \Big(|H_{\frac12 1}|^2 -|H_{-\frac12 -1}|^2\Big)
  + \frac 32 \cos^2\theta_V \Big(|H_{\frac12 0}|^2 -|H_{-\frac12 0}|^2 \Big)}
   {\frac 34 \sin^2\theta_V \Big(|H_{\frac12 1}|^2 +|H_{-\frac12 -1}|^2\Big)
   +\frac 32 \cos^2\theta_V \Big(|H_{\frac12 0}|^2 +|H_{-\frac12 0}|^2 \Big)} \,. 
\en
When averaged over $\cos\theta_V$ (one has to integrate the numerator and
denominator separately) one has  
\be
P_{B_2}=   P_{\Sigma_c^{++}}=
    (|\hat H_{\frac12 1}|^2 -|\hat H_{-\frac12 -1}|^2 )
+ (|\hat H_{\frac12 0}|^2 -|\hat H_{-\frac12 0}|^2 )
={\cal F}^P_T + {\cal F}^P_L \,. 
\en
As mentioned before the polarization of the $\Sigma_c^{++}$ is not 
measurable in its strong decay. However, the $\Sigma_c^{++}$  
transfers its polarization to the $\Lambda_c^+$ in the (strong) decay  
$\Sigma_c^{++} \to \Lambda_c^+\ \pi^+$ where the polarization transfer 
depends on $\cos\theta_B$. The average longitudinal polarization of the 
$\Lambda_c^+$ can be calculated to be (again we average over $\cos\theta_V$)
\be
     P_{\Lambda_c^+}(\theta_B)=P_{\Sigma_c^{++}}
      \,\cos\theta_B  \,. 
\en
As in Eq.~(\ref{reversal}) one has a sign reversal for the $\Xi_c^+$ 
in the decay chain $\Xi_c^{'+} + \Xi_c^+ + \gamma$.
\item \quad $1/2^+ \to 3/2^+ + 0^-$ \\[10pt]
In Table~\ref{tablenum3} we list the rates and branching fractions for the
four decays in this class.   

\begin{table}[ht]
\caption
    {\label{tablenum3}Cabibbo-favored factorizing nonleptonic two-body decays
      of double heavy charm baryons
      induced by the quark level $c \to s;d \to u$ transitions for the
      cases $1/2^+ \to 3/2^+ +0^-$.}
   \vspace{0.2cm}
\begin{center}
\begin{tabular}{llccc}
  \hline
  \hline
\phantom{$1/2^+ \to 3/2^+ +0^-$}&\phantom{$\hat{\hat{I}_3}$} 
& \,\,\,$ \Gamma$ [$10^{-13}$ GeV] \,\,\, 
& \,\,\, $\cal{B}$ $[\%]$ \,\,\, \\ 
      \hline
$1/2^+ \to 3/2^+ + 0^-$ & 
$\Xi_{cc}^{++} \to \Sigma_{c}^{*++} + \bar K^0$ 
& 0.06 & 0.25 
\\[0.5ex]
  \phantom{$1/2^+ \to 3/2^+ +0^-$}&
$\Xi_{cc}^{++} \to \Xi_{c}^{*+} + \pi^+$ 
& 0.16 & 0.63 
\\[0.5ex]
  \phantom{$1/2^+ \to 1/2^+ + 0^-$}&
$\Omega_{cc}^{+} \to  \Xi_{c}^{*\,+} +\bar K^0$ 
& 0.03 & 0.10  
 \\[0.5ex]
 \phantom{$1/2^+ \to 3/2^+ + 0^-$} &
$\Omega_{cc}^{+} \to \Omega_{c}^{*0}+\pi^+$ 
& 0.31 & 1.00 
\\[0.5ex]
\hline
\hline
\end{tabular} 
\end{center}
\end{table}    
Contrary to the $1/2^+ \to 1/2^+ + 0^-$ case the baryon side angular decay  
distribution of $\Sigma_c^{++}(2520;3/2^+) \to \Lambda_c^+ \pi^+$ 
now shows a $\theta_B$ dependence given by
\be
W(\theta_B)= (1-\tfrac34 \sin^2\theta_B)\,,
\en
i.e. there is a pronounced dip of the angular decay distribution at
$\theta_B=90^\circ$.

In the constituent quark model the vector transition 
$1/2^+ \to 3/2^+$ is conserved, i.e. the vector current 
helicity amplitude $H^V_{\pm1/2\,t}=0$ vanishes for the transition 
$\Xi_{cc}^{++} \to \Sigma_c^{*++}$. This implies that the final  
$3/2^+$ state has no polarization structure and therefore there 
is no polarization transfer to the  $\Lambda_c^+$ in the second stage 
decay $\Sigma_c^{++}(2520;3/2^+) \to \Lambda_c^+ \pi^+$. The same statement 
holds true for the other three $1/2^+ \to 3/2^+ + 0^-$ decays. 

\vspace*{.2cm}     
\item \quad $1/2^+ \to 3/2^+ + 1^-$ 
\vspace*{.2cm}     

In Table~\ref{tablenum5} we list the rates, branching fractions and the 
three  polarization parameters ${\cal F}^{P}_L$,  
${\cal F}^{P}_T$ and ${\cal F}^{P'}_T$ needed to describe the longitudinal 
polarization of the daughter baryon $P_{B_2}$ for the four decays in this 
class.
       
\begin{table}[ht]
\caption
{\label{tablenum5}Cabibbo favored factorizing nonleptonic two-body decays 
of double heavy charm baryons induced by the quark level 
$c \to s;d \to u$ transitions for the case $1/2^+ \to 3/2^+ + 1^-$.} 
\vspace{0.2cm}
\begin{center}
\begin{tabular}{llccccc}
  \hline
  \hline
\phantom{$1/2^+ \to 1/2^+ +0^-$}&\phantom{$\hat{\hat{I}_3}$} 
& \,\,\, $\Gamma$ [$10^{-13}$ GeV] \,\,\, 
& \,\,\, $\cal{B}$ $[\%]$ \,\,\, 
& \,\,\, ${\cal F}^P_L$ \,\,\, 
& \,\,\, ${\cal F}^P_T$ \,\,\, 
& \,\,\, ${\cal F}^{'P}_T$ \,\,\, \\ 
      \hline
$1/2^+ \to 3/2^+ + 1^-$ \qquad \qquad &
$\Xi_{cc}^{++} \to \Sigma_{c}^{*++} + \bar K^{*0}$
      \qquad & 0.42 & 1.62 & -0.01 & -0.10 & -0.31 
\\[0.5ex]
      \phantom{$1/2^+ \to 3/2^+ +1^-$}&
$\Xi_{cc}^{++} \to \Xi_{c}^{*+} + \rho^+$
       & 1.15 & 4.48 & -0.01 & -0.08 & -0.24  
\\[0.5ex]
      \phantom{$1/2^+ \to 3/2^+ + 1^-$}&$\Omega_{cc}^{+} \to \Xi_{c}^{*+}
      +\bar K^{*0}$
    &  0.21 & 0.67 & -0.01 & -0.10 & -0.30 
\\[0.5ex]
\phantom{$1/2^+ \to 3/2^+ +1^-$}&$\Omega_{cc}^{+} \to \Omega_{c}^{*0} +
 \rho^+$ 
& 2.23 & 7.11 & -0.01 & -0.08 & -0.24 
\\[0.5ex]  
\hline
\hline
\end{tabular} 
\end{center}
\end{table}

The threefold joint angular decay distribution can be obtained 
from the first row of Eq.(\ref{angdist2}) by replacing the 
Wigner $d^{1/2}$--function by the corresponding spin 3/2 Wigner 
$d^{3/2}$--function. 
Again, one has to observe the  angular momentum  
constraint $|\lambda_2 -\lambda_M|\le 1/2$. The threefold angular decay 
distribution reads
\bea
\label{angdist3}
           W(\theta_V,\theta_B,\chi)&=&\frac 32 \cos^2\theta_V
           (1- \frac 34 \sin^2\theta_B)(|\hat H_{\frac12 0}|^2+
           |\hat H_{-\frac12 0}|^2) \nn
           &+& \frac 34 \sin^2\theta_V\Big((1- \frac 34 \sin^2\theta_B)
           (|\hat H_{\frac12 1}|^2+|\hat H_{-\frac12 -1}|^2)
           +\frac 34 \sin^2\theta_B(|\hat H_{\frac32 1}|^2
           +|\hat H_{-\frac32 -1}|^2)\Big) \nn
           &+& \frac {3}{8}  \Big(\sqrt{\frac 32}\sin 2\theta_V \sin 2\theta_B
           \Big(\cos \chi \, 
           {\rm Re}(\hat H_{\frac32 1} \hat H^\ast_{\frac12 0}+
           \hat H_{-\frac32 -1} \hat H^\ast_{-\frac12 0}) \nn 
           &&\qquad\qquad\qquad\qquad\quad\hspace{.15cm} +
           \sin\chi \, 
           {\rm Im} (\hat H_{\frac32 1} \hat H^\ast_{\frac12 0}-
           \hat H_{-\frac32 -1} \hat H^\ast_{-\frac12 0})
           \Big) \nn
           &+& 3 \sqrt{3}\sin^2 \theta_V \sin^2\theta_B 
           \Big(\cos2 \chi \, {\rm Re}(\hat H_{\frac32 1}
           \hat H^\ast_{-\frac12 -1}+ \hat H_{-\frac32 -1}
           \hat H^\ast_{\frac12 1}) \nn
           &&\qquad\qquad\qquad\qquad +\sin2\chi \, 
           {\rm Im}(\hat H_{\frac32 1}
           \hat H^\ast_{-\frac12 -1}- \hat H_{-\frac32 -1}
           \hat H^\ast_{\frac12 1})
           \Big)  \,. 
\ena
The angular decay distribution can be seen to integrate to $2\pi$. 
It is apparent that there is a rich angular structure in the 
angular decay distribution~(\ref{angdist3}).

In Eq.(\ref{angdist3}) we have also included the $T$--odd 
contributions proportional to ${\rm Im}(\hat H_{\frac32 1} 
\hat H^\ast_{\frac12 0}-\hat H_{-\frac32 -1} \hat H^\ast_{-\frac12 0})$  
and ${\rm Im}(\hat H_{\frac32 1} \hat H^\ast_{-\frac12 -1} 
- \hat H_{-\frac32 -1} 
\hat H^\ast_{\frac12 1})$ even though these contributions vanish in 
our model calculation because our helicity amplitudes are relatively real. 
The angular coefficients that multiply the $T$--odd contributions can be 
seen to involve $T$--odd triple products. For example, one has
\be
           \sin 2\theta_V \sin 2\theta_B \sin\chi= 4(\hat p_{\Sigma^{++}_c}
           \cdot \hat p_{\Lambda_c})(\hat p_{K^{*0}}\cdot \hat p_{k^-})
           (\hat p_{\Sigma^{++}_c}\times
           \hat p_{\Lambda_c})\cdot \hat p_{k^-}
\en
where the hatted three-momenta are normalized to 1. 
The $T$--odd can be fed by either final state interactions or by  
$CP$--violating interactions. It would be interesting to experimentally 
check on the existence of such triple-product correlations.

We define polarization parameters
that describe the angular decay distribution where we also include
the numerical values for the parameters for the cascade decay
$\Xi_{cc}^{++} \to \Sigma_{c}^{*++}(\to \Lambda_c^+ + \pi^+) +
\bar K^{*0}(\to K^- + \pi^+)$. One has
\bea\label{coeff}
{\cal F}_L&=&|\hat H_{\frac12 0}|^2+|\hat H_{-\frac12 0}|^2
               = 0.40\,, \nn
{\cal F}_T &=& |\hat H_{\frac12 1}|^2+|\hat H_{-\frac12 -1}|^2
               = 0.16\,, \nn
{\cal F}^{'}_T &=& |\hat H_{\frac32 1}|^2+|\hat H_{-\frac32 -1}|^2
               = 0.45\,, \nn
{\cal F}^{P }_L&=&|\hat H_{\frac12 0}|^2-|\hat H_{-\frac12 0}|^2
               = -0.01\,, \nn
{\cal F}^{P }_T &=& |\hat H_{\frac12 1}|^2-|\hat H_{-\frac12 -1}|^2
               = -0.10\,, \nn
{\cal F}^{P '}_T &=& |\hat H_{\frac32 1}|^2-|\hat H_{-\frac32 -1}|^2
               = -0.31\,, \nn
\gamma &=& {\rm Re}(\hat H_{\frac32 1} \hat H^\ast_{\frac12 0}+
               \hat H_{-\frac32 -1} \hat H^\ast_{-\frac12 0})
               = 0.39\,, \nn
\gamma'&=& {\rm Re}(\hat H_{\frac32 1}
           \hat H^\ast_{-\frac12 -1}+ \hat H_{-\frac32 -1}
           \hat H^\ast_{\frac12 1})= 0.19\,. 
\ena

By integrating over the respective pairs of angles 
one obtains the single angle decay distributions
\bea
\frac{1}{2\pi}\,W(\theta_V)&=& \frac 32 \cos^2\theta_V {\cal F}_L 
+ \frac 34 \sin^2 \theta_V({\cal F}_T+{\cal F}^{'}_T) \,,\nn
\frac{1}{2\pi}\,W(\theta_B)&=& (1- \frac 34 \sin^2\theta_B )
({\cal F}_L+{\cal F}_T)
+ \frac 34 \sin^2\theta_B {\cal F}^{'}_T \,,\nn
W(\chi)&=& 1+2\sqrt{3}\,\gamma'\, \cos2\chi  \,. 
\ena
The contribution of the remaining azimuthal asymmetry parameter
$\gamma$ can be obtained by folding the angular decay distribution 
with $\cos\theta_V\,\cos\theta_B$. The numerical values of the 
angular coefficients are listed in Eq.~(\ref{coeff}). 
           
The polarization transfer to the $\Lambda_c$ in the strong 
decay $\Sigma_c^{*\,++} \to \Lambda_c^+ +\pi^+$ is given by 
\bea
P_{\Lambda_c^+}(\theta_B)&=&\Big( 
(|\hat H_{\frac12 0}|^2-|\hat H_{-\frac12 0}|^2+|\hat H_{\frac12 1}|^2-
|\hat H_{-\frac12 -1}|^2)
(1-\frac 34\sin^2\theta_B) \nn
&+&(|\hat H_{\frac32 1}|^2-|\hat H_{-\frac32 -1}|^2)
\frac 34 \sin^2\theta_B\Big)
\,\cos\theta_B \,.
\ena
As in~Eq.(\ref{reversal}) there is a reversal in sign in the 
polarization transfer to the stage 2 charm baryons $\Xi_c^+$ and 
$\Omega_c^0$ in the last three decays of Table~\ref{tablenum5} 
since the stage 2 decays $\Xi_{c}^{*+} \to \Xi_{c}^{+} + \gamma$ 
and $\Omega_{c}^{*0} \to \Omega_{c}^{*0}+ \gamma$ are $ \sim100\,\%$ 
one-photon decays.

\end{itemize}       

\section{Summary and conclusion}

   We have cataloged all Cabibbo-favored semileptonic and nonleptonic 
   two-body decays of the three double heavy charm baryon states
   $\Xi_{cc}^{++},\,\Xi_{cc}^{+}$ and $\Omega_{cc}^+$ where the nonleptonic
   two-body decays are into ground state mesons and baryons. For the
   semileptonic decays we have calculated rates, branching ratios and
   helicity fractions of the $W^-_{\rm off-shell}$ using transition form
   factors calculated in our CCQM quark model. For the nonleptonic
   decays we have analyzed the topology structure of their various two-body
   decays in terms of the two $W$--emission (external and internal) or tree
   topologies
   and the three $W$--exchange topologies. We have identified two groups
   of decays $\Xi_{cc}^{++} \to \Sigma_{c}^{++\,(*)} + \bar K^{0 (*)}$ and
   $\Omega_{cc}^{+} \to \Omega_c^{0\,(*)} + \pi^+(\rho^+)$ which proceed by
   $W$--emission alone and are thus theoretically favored since 
   there is no contamination from $W$--exchange contributions. The
  $W$--exchange contributions to the decays 
  $\Xi_{cc}^{++} \to \Xi_{c}^{\prime(*)+} + \pi^+(\rho^+)$ and
   $ \Omega_{cc}^{+} \to \Xi_{c}^{\prime (*)+} +\bar {K}^{(*)0}$ vanish in the
   $SU(3)$ limit as a consequence of the K\"orner-Pati-Woo theorem
   Using again
   transition amplitudes from our CCQM quark model for the latter decays
   we have calculated rates, branching ratios
   and angular coefficients that characterize the angular decay distributions
   of the one-sided or two-sided cascade decays of the above two classes of
   decays. The angular decay
   distributions involving the $1/2^+ \to 1/2^+$ baryon transitions are
   markedly different from those of the $1/2^+ \to 3/2^+$ transitions.
   In particular, in the $1/2^+ \to 1/2^+ 1^-$ cascade decay
   $\Xi_{cc}^{++} \to \Sigma_{c}^{++}(2455;1/2^+) + \bar K^{*\,0}$ there are no
   azimuthal correlations between the two planes formed by the second stage
   decays $\Sigma_{c}^{++}(2455;1/2^+) \to \Lambda_c^+ \pi^+$
   and $ \bar K^{*0} \to K^-  \pi^+)$
   whereas the two decay planes become azimuthally correlated in the
   $1/2^+ \to 3/2^+ + 1^-$ cascade decay
   $\Xi_{cc}^{++} \to \Sigma_{c}^{*++}(2520;3/2^+) + \bar K^{*0}$. Another
   discriminating feature of these two possible decay paths
   is that there is  pronounced dip in the $\cos\theta$ distribution in the
   latter case.
   
   Any of the two-body nonleptonic decays of the $\Xi^+_{cc}$ and
   $\Omega_{cc}^{+}$ listed in Table~\ref{topol} could be explored in the
   search for the two missing double heavy charm baryon states. 
   If one takes the
   discovery channels of the $\Xi^{++}_{cc}$ as a guide the decays
    $\Xi_{cc}^{+} \to \Lambda_c^{+}(\Sigma_c^{(*)+})
   +\bar K^{(*)0}$ and $\Xi^+_{cc} \to \Xi_{c}^{0\,(\prime ,*)}+ \pi^+(\rho^+)$
   would be good candidates for the discovery of the $\Xi_{cc}^{+}$ while the
   $\Omega_{cc}^{+}$ should be searched for in the decays
   $\Omega_{cc}^{+} \to \Xi_{c}^{+\,(\prime ,*)} 
   +\bar {K}^{(*)0}$ or $\Omega_{cc}^{+} \to \Omega_c^{0\,(*)} 
   + \pi^+(\rho^+)$. In this paper we have provided first predictions for the
   branching ratios of the decays 
   $\Xi_{cc}^{++} \to \Sigma_{c}^{++\,(*)} 
   + \bar K^{0 (*)}$, $\Xi_{cc}^{++} \to \Xi_{c}^{+\,(\prime,* )} 
   + \pi^+(\rho^+)$, $\Omega_{cc}^{+} \to \Xi_{c}^{+\,(\prime ,*)}  
   + \bar {K}^{(*)0}$ and $\Omega_{cc}^{+} \to \Omega_c^{0\,(*)} 
   + \pi^+(\rho^+)$. In a follow-up paper we plan to also calculate the
   $W$--exchange contribution to the Cabibbo favored nonleptonic double charm
   baryon decays with predictions for the
   remaining decays of Table~\ref{topol} not treated in this paper. 
   This includes a calculation of the recently observed decay 
   $\Xi_{cc}^{++} \to \Xi_c^+ +\pi^+$~\cite{Aaij:2018gfl}. 

\begin{acknowledgments}

This work was funded by
the Carl Zeiss Foundation under Project ``Kepler Center f\"ur Astro- und
Teilchenphysik: Hochsensitive Nachweistechnik zur Erforschung des
unsichtbaren Universums (Gz: 0653-2.8/581/2)'', 
by ``Verbundprojekt 05A2017 - CRESST-XENON: Direkte Suche nach Dunkler 
Materie mit XENON1T/nT und CRESST-III. Teilprojekt 1'' 
(F\"orderkennzeichen 05A17VTA)'', 
by CONICYT (Chile) under
Grants No. PIA/Basal FB0821 and by FONDECYT (Chile) under Grant No. 1191103.
M.A.I.\ acknowledges the support from the PRISMA Cluster of Excellence
(Mainz Uni.). M.A.I. and J.G.K. thank the Heisenberg-Landau Grant for
partial support.

\end{acknowledgments}

\end{document}